\documentclass[prl,aps,twocolumn,groupedaddress,superscriptaddress,floatfix,showpacs]{revtex4}
\usepackage{epsfig}
\usepackage{mathrsfs}
\usepackage{amsmath}
\usepackage{textcmds}
\allowdisplaybreaks[1]
\newcommand{\be}{\begin{equation}}
\newcommand{\ee}{\end{equation}}
\newcommand{\bea}{\begin{eqnarray}}
\newcommand{\eea}{\end{eqnarray}}
\newcommand{\ket}[1]{\left|#1\right\rangle}

\newcommand{\braket}[2]{\left\langle #1 | #2 \right\rangle}

\newcommand{\bc}{\begin{center}}
\newcommand{\ec}{\end{center}}

\renewcommand{\(}{\left(}
\renewcommand{\)}{\right)}

\newcommand{\forget}[1]{}

\newcommand{\re}{{\rm e}}

\newcommand{\ri}{{\rm i\,}}
\begin{document}
\title{The Vortex Phase Qubit:  Generating Arbitrary, Counter-Rotating, Coherent Superpositions in Bose-Einstein Condensates via Optical Angular Momentum Beams}
\author{Kishore T. Kapale}
\email{Kishor.T.Kapale@jpl.nasa.gov}
\affiliation{Quantum Computing Technologies Group, Jet Propulsion Laboratory,
California Institute of Technology, Mail Stop 126-347, 4800 Oak Grove Drive, 
Pasadena, California 91109-8099}
\author{Jonathan P. Dowling}
\affiliation{Hearne Institute for Theoretical Physics, Department of Physics \& Astronomy,
Louisiana State University,
Baton Rouge, Louisiana 70803-4001}
\affiliation{Institute for Quantum Studies, Department of Physics, Texas A\&M  University, College Station, Texas 77843}
\begin{abstract}
We propose a scheme for generation of arbitrary coherent superposition of vortex states in Bose-Einstein condensates (BEC) using the orbital angular momentum (OAM) states of light. We devise a scheme to generate coherent superpositions of two counter-rotating OAM states of light using known experimental techniques.  We show that a specially designed Raman scheme allows transfer of the optical vortex superposition state onto an initially non-rotating BEC. This creates an arbitrary and coherent superposition of a vortex and anti-vortex pair in the BEC. The ideas presented here could be extended to generate entangled vortex states, design memories for the OAM states of light, and perform other quantum information tasks. Applications to inertial sensing are also discussed.
\end{abstract}
\maketitle

\forget{\it Introduction.---}
Generation and manipulation of macroscopic superpositions as well as entangled states is of paramount interest to the field of
quantum information~\cite{Nielsen:2000}. In this regard, Bose-Einstein condensates (BEC)~\cite{Anderson:1995, Bradley:1995,Mewes:1996} come across as ideal candidates. BECs
correspond to highly-coherent macroscopic ground-states of the confining potentials. Moreover, vortex states of BECs, which are topological states with special phase structure, have been realized experimentally~\cite{Abo-Shaeer:2001}. Stirring a BEC cloud with laser beams leads to the nucleation of vortex lattices in the BEC. These vortex
states are fairly stable and could be candidates for qubits in quantum information processors, if appropriate means to manipulate them are developed.

In an entirely different area of optical physics, tremendous progress has been made in creation~\cite{Heckenberg:1992,Agarwal:1997OAM, Arlt:1998, Sueda:2004}, manipulation~\cite{Akamatsu:2003}, detection~\cite{Molina-Terriza:2002,Leach:2002,Bigelow:2004},  and application~\cite{Grier:2003} of the orbital angular momentum (OAM) states of light.  The OAM states have a corkscrew type helical
phase structure.  To illustrate, an OAM state with angular momentum
$\hbar\ell$  has $|\ell|$  azimuthal phase singularities across a cut taken in the beam path. The sign of $\ell$ corresponds to the sense of rotation of the phase fronts around the beam axis. Each photon in the OAM beam carries an orbital angular momentum of $\hbar\ell$. \forget{In principle, there is no upper limit to the angular momentum value that can be imparted to a light beam. Therefore, OAM states of light themselves are very good candidates for quantum information processing and in particular for quantum cryptography~\cite{Spedalieri:2004QKD},  as they provide essentially infinite Hilbert space to work with thus increasing the security of the quantum-key distribution protocols tremendously.}  The quantum nature of these OAM states has been demonstrated recently by showing that a photon pair created in parametric down-conversion process is entangled in the orbital angular momentum space along with the usual polarization entanglement~\cite{Mair:2001}.

Excitation of vortices in BECs, using the optical vortex beams, has been proposed recently~\cite{Marzlin:1997,Nandi:2004}. In this letter we introduce a scheme for creation of macroscopic superpositions of BEC vortex states through transfer of angular momentum of light from specially prepared OAM state superpositions. \forget{First, we design a scheme for creation of superposition of OAM
states of light, then we discuss the mechanism to transfer the superposition state to the non-rotating BECs to achieve macroscopic superposition of vortex states.}

\forget{\it Superposition of optical vortices.---}
Generation of the superposition of gaussian beams with OAM states of light has been demonstrated~\cite{Vaziri:2002}. Our interest, however, lies in creating an arbitrary superposition of two counter-rotating optical vortices. 

The OAM states of light have unique amplitude and phase structures. To illustrate, monochromatic OAM beams have an azimuthal phase dependence of the type $\exp(\ri \ell \phi)$.  
Laguerre-Gaussian (LG) laser modes are an example of such OAM states~\cite{Allen:1992}. The normalized LG mode at the beam waist $(z=0)$ and beam size
$w_0$ at the waist is given in 
cylindrical coordinates ($\rho,\phi,z$) by
\begin{align}
\label{Eq:LG}
\mbox{LG}_{p}^l(\rho, \phi)& = \sqrt{\frac{2 p !}{\pi(|\ell|+p)!}} \frac{1}{w_0} 
\left(\frac{\sqrt{2}\rho}{w_0}\right)^{|\ell|} L^{|\ell|}_p\left(\frac{2 \rho^2}{w_0^2}\right)\nonumber \\
&\qquad\exp{(-\rho^2/w_0^2)}\exp{(\ri \ell \phi)}
\end{align}
where $L^l_p(\rho)$ are the associated Laguerre polynomials,
\begin{equation}
L^{|\ell|}_p(\rho)=\sum_{m=0}^p (-1)^m \frac{(|\ell|+p)!}{(p-m)!(|\ell|+m)!m!}\rho^m\,;
\end{equation}
$w_0$ is the beam width, $p$ is the number of non-axial radial nodes of the mode, and the index $\ell$, referred to as
the winding number, which describes the helical structure of the wave front around a phase dislocation.
For further discussion we consider only pure LG modes with charge $\ell$ and $p=0$, we denote such a state of the
light field by $\ket{\ell}$ such that
$
\braket{\mathbf r}{\ell} = \mbox{LG}^{\ell}_0(\rho, \phi).
$
Thus it can be easily seen that the states $\ket{+\ell}$ and $\ket{-\ell}$, with $\ell$ being a whole number, differ only in the sense of the winding of the phase either clockwise or counter-clockwise. Our aim is to create a general superposition of the OAM states of light of the kind:
$(a_{+}\ket{\ell}+a_{-}\ket{-\ell})$, with $|a_{+}|^2 + |a_{-}|^2=1$.
It is well known that creation of superposition of OAM states of the kind $\sum_{\ell} c_{\ell} \ket{\ell}$ 
is a fairly straightforward procedure by using computer generated holographs~\cite{Arlt:1998} or phase plates~\cite{Sueda:2004}.
Moreover, a sorter of these OAM states has also been demonstrated~\cite{Leach:2002} that can distinguish and separate different OAM components. Thus, by using a mixed OAM-state generator and a OAM-sorter in conjunction one can easily obtain a pure OAM state $\ket{\ell}$.  

\begin{figure}[ht]
\includegraphics[scale=0.33]{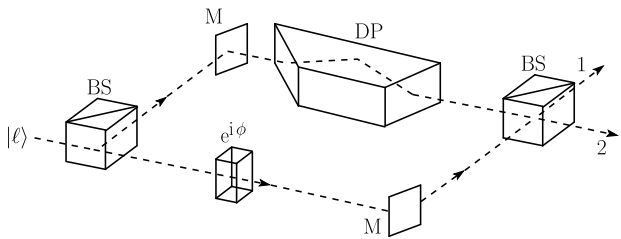}
\caption{\label{Fig:OAMS} Scheme for creation of superposition of the OAM states.
The first beam splitter (BS) at the input is a $|\tilde{r}|^2/|\tilde{t}|^2$ beam splitter as $\tilde{r}$ and $\ri \tilde{t}$ are the reflection and transmission amplitudes. The second beam-splitter is taken to be 50:50, with the mirrors (M) perfectly reflecting. The dove prism (DP) transforms a right handed coordinate-system into a left handed one; thus it performs the operation $\ket{\ell} \rightarrow \ket{-\ell}$. Choosing $\phi=\pi$, and discarding the output at the output port 2, one obtains $(\tilde{t}\ket{+\ell}+\tilde{r}\ket{-\ell})$ at output port 1.}
\end{figure}

We note that dove prisms can be used to change the handedness of light beams passing through them~\cite{Padgett:1999}. Consequently, the sense of the phase winding of an LG beam would be reversed as it passes through a dove prism. Using this, we devise a Mach-Zender type configuration as shown in Fig.~\ref{Fig:OAMS} to generate a general superposition $(\tilde{t}\ket{\ell}+\tilde{r}\ket{-\ell})$ at one of the output ports of the interferometer. The first beam splitter is taken to be a special beam splitter with the ratio of $\tilde{\ri t}:\tilde{r}$ for the transmitted and the reflected amplitudes at its output ports. The second beam splitter is a usual 50:50 beam splitter. 
The operation of the Mach-Zender configuration of Fig.~\ref{Fig:OAMS} can be described through the matrix representation of beam splitter operation~\cite{Zeilinger:1981}, such that the initial state  $(\ket{\ell}, \quad 0)^{\rm T}$
transforms into
\forget{
\begin{equation}
\frac{1}{\sqrt{2}}\(
\begin{matrix}
 \tilde{r} \ket{-\ell} - \re^{\ri \phi}\, \tilde{t}\ket{\ell}  \\
 \ri \tilde{r} \ket{-\ell} + \ri \tilde{t} \re^{\ri \phi}\ket{\ell}
\end{matrix}
\)
\end{equation}
Thus, with the choice of the phase $\phi=\pi$ we obtain the state}
\begin{equation}
\frac{1}{\sqrt{2}}\(\begin{matrix} \tilde{t}\ket{\ell}  + \tilde{r}\ket{-\ell} \\   
\ri(\tilde{r}\ket{-\ell}  - \tilde{t}\ket{\ell}) \end{matrix}\)
\end{equation}
at the exit ports 1 and 2 of the Mach Zender Interferometer, respectively with the choice of $\phi=\pi$.
Thus, by ignoring port 2 and renormalizing the state from port 1 we obtain the required superposition state $\tilde{t}\ket{\ell}  + \tilde{r}\ket{-\ell}$, as $|\tilde{r}|^2 + |\tilde{t}|^2 =1$. In the following we present our scheme for transfering this optical vortex state superposition to BEC vortex  superpositions.

Highly detuned optical fields in a Raman configuration have been used to coherently manipulate 
and create various superpositions of different atomic levels~\cite{Niu:2004}. Similar optical manipulation techniques exist to couple BEC clouds in different internal states. Moreover, the OAM states of light have also been shown to be useful for excitation of vortices in BEC through time-dependent linearily varying two-photon detuning~\cite{Marzlin:1997} and through the STIRAP~\cite{Bergmann:1998} type scheme~\cite{Nandi:2004}. In the following we discuss a Raman type scheme to generate superposition of vortex states in BEC.

\begin{figure}[ht]
\centerline{\includegraphics[width=0.8\columnwidth]{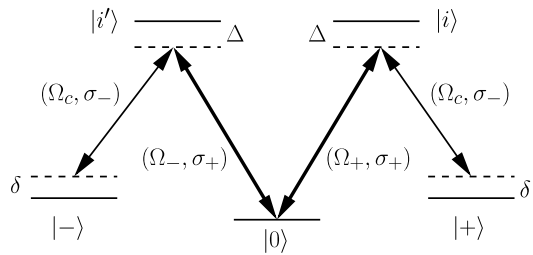}}
\caption{\label{Fig:LevelScheme}The level scheme for generation of vortex state superposition.  A non-rotating state $\ket{0}$ is coupled to the vortex states $\ket{+}$ and $\ket{-}$ through the optical vortex field components providing coupling strengths $\Omega_{+}$, $\Omega_{-}$ and a strong drive field with Rabi frequency $\Omega_c$. The optical vortex beam is  $\sigma_{+}$ polarized, whereas the drive field is $\sigma_{-}$ polarized such the hyperfine quantum number $m_F$ is the same for the components $\ket{0}$, $\ket{\pm}$. The vorticity of the $\ket{\pm}$ states is $\pm \ell$, respectively}
\end{figure}

The level scheme for our model is  depicted in Fig.~\ref{Fig:LevelScheme}. Initially nonrotating state $\ket{0}$ is coupled optically via two Raman type configurations of the external fields, $(\Omega_{+}, \Omega_c)$ and $(\Omega_{-}, \Omega_{c})$,  through  internal states $\ket{i}$ and $\ket{i'}$. The polarizations of the optical fields are taken as shown in the figure; thus the internal quantum numbers of the final states $\ket{+}$ and $\ket{-}$ are the same. The Rabi frequencies  $\Omega_{+}$ and $\Omega_{-}$ are due to the coupling of the BEC cloud with the two counter-rotating components of the special optical vortex state generated through the configuration discussed in Fig.~\ref{Fig:OAMS}

\begin{figure}[ht]
\includegraphics[width=0.8\columnwidth]{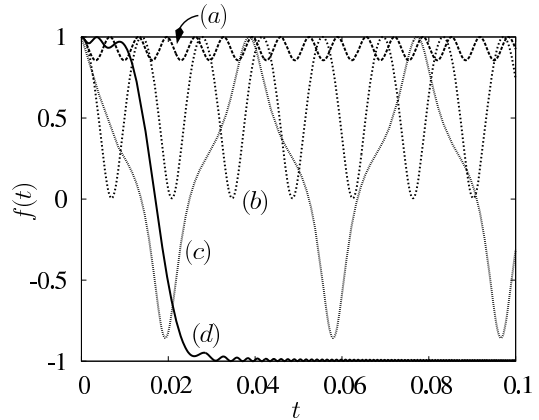}
\caption{\label{Fig:TF} Transfer function $f(t)$ for various detunings. (a) $\delta = 0$, (b) $\delta=900$ , (c) $\delta=380$,   (d) linearly varying detuning $\delta(t) =  3000-400^2 t$. All detunings are given in Hz. The complete population transfer to the vortex state, corresponding to $f(t) =-1$ for sufficiently large $t$, is possible only with time-dependent detuning.}
\end{figure}

\begin{figure*}[ht]
\includegraphics[width=1.0\textwidth]{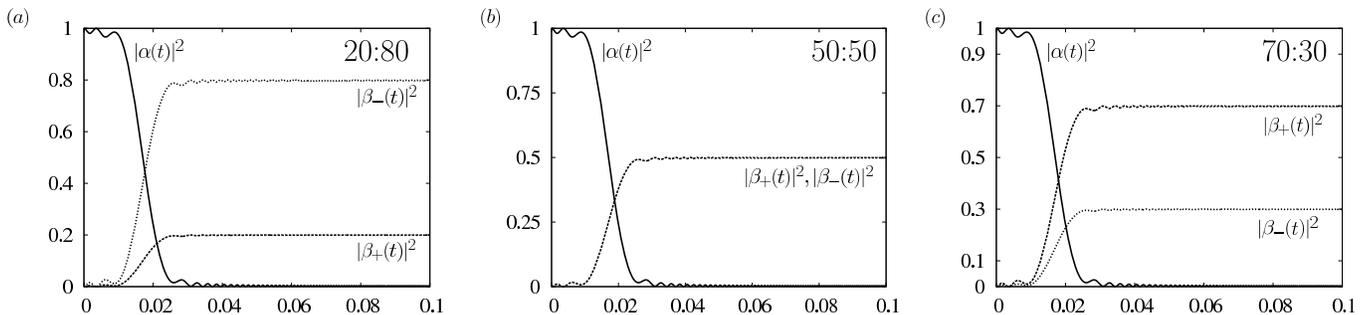}
\caption{\label{Fig:Pop} Various superpositions of the vortex states.  The steady state population ratios are give in the notation $|\beta_{+}|^2:|\beta_{-}|^2$  The vertical axis on all the plots correspond to the populations of various states and the horizontal axis is the time measured in seconds.}
\end{figure*}

Noting that the optical fields are highly detuned, the intermediate states ($\ket{i}$ and $\ket{i'}$)  are sparingly populated, Thus,  they can be adiabatically eliminated from the equations. Within this adiabatic approximation we can write a modified set of Ginzburg-Pitaevskii-Gross (GPG)~\cite{Edwards:1996} equations for the multi-component BEC trapped by the cigar-shaped trapping potential $\mathscr{V}=(m/2)\,(\omega_{\perp}^2 r^2+ \omega_z z^2)$, where $\omega_{\perp}$ and $\omega_z$ are the transverse and longitudinal trapping frequencies respectively. Also $m$ is the mass of the individual atoms in the BEC cloud. We note that $r = \sqrt{x^2 +y^2}$ is the transverse radial coordinate and $\phi$ would be the azimuthal angle in the $x$-$y$ plane that would be required later to describe the phase structure of the rotating BECs. The configuration space representations of the states $\ket{0}, \ket{+}$ and $\ket{-}$ shown in Fig.~\ref{Fig:LevelScheme} are taken to be $\Psi_{0}, \Psi_{+}$ and $\Psi_{-}$ respectively.  Thus, including the Raman-type optical couplings arising from the adiabatic elimination of the intermediate state, we arrive at the modified GPG equations for the three relevant components of the BEC cloud
\begin{align}
\ri \hbar \dot{\Psi}_{0} &=\mathscr{H}\Psi_0 
                       + \frac{\hbar}{\Delta}\(\sum_{i=\pm} |\Omega_i|^2 \Psi_{0}
                       + \sum_{i=\pm} \Omega_{i}^* \Omega_c e^{-i \Delta t}\Psi_{i}\) \nonumber \\
\ri \hbar \dot{\Psi}_{\pm} &= \mathscr{H}\Psi_{\pm} 
		       + \frac{\hbar}{\Delta}\(|\Omega_c|^2\Psi_{\pm}
		       +\Omega_{\pm}\Omega_c^* e^{i \Delta t}\Psi_0\)
\label{Eq:psidot}
\end{align}
where $\mathscr{H} = \mathscr{T}+\mathscr{V}-\mu +  \eta (  |\Psi_0|^2 +  |\Psi_+|^2 +  |\Psi_-|^2 )$ is the self-energy operator which includes the kinetic ($\mathscr{T}$), potential ($\mathscr{V}$), and the interaction energy operators for the BEC states. Here $\eta = 4 \pi \hbar a_{\rm sc} N/m$ signifies the strength of the inter-particle interactions of the $N$-particle BEC cloud, through the scattering cross-section $a_{\rm sc}$.  Using the LG beam mode function Eq.~\eqref{Eq:LG}, the Rabi frequencies corresponding to the coupling between the optical vortex beam superposition $a_{+}\ket{\ell}+ a_{-}\ket{-\ell}$ and the BEC cloud can be written as
\begin{equation}
\label{Eq:Omegapm}
\Omega_{\pm} ({\mathbf r}) = a_{\pm} \Omega_0\,  \re^{-r^2/w^2} ( {\sqrt{2} r}/{w})^{|\ell|} \re^{\pm\ri \ell \phi } \re^{\ri k z} 
\end{equation} 
with $\Omega_0$ is the usual atom-field interaction Rabi frequency. The amplitudes $a_{\pm}$ are respectively $\tilde{t}$ and $\tilde{r}$ corresponding to the first beam splitter transmission and reflection amplitudes of the configuration shown in Fig.~\ref{Fig:OAMS}.
With the size of the condensate chosen to be much smaller than the Laguerre Gaussian beam waist, the exponential dependence on the radial coordinate in Eq.~\eqref{Eq:Omegapm} can be ignored.
Now we make an ansatz that the topological structure of the states $\ket{0}$, $\ket{\pm}$ are given by
\begin{subequations}
\label{Eq:topology}
\begin{align}
\Psi_{0}({\mathbf r}, t) &= \alpha(t) \exp[\ri(\mu/\hbar - \kappa)t] \,\psi_g({\mathbf r})\,,\nonumber \\
\Psi_{\pm}({\mathbf r}, t) &= \beta_{\pm}(t)\exp[\ri(\delta+ \mu/\hbar - \kappa)t] \, \psi_{v\pm}({\mathbf r})\,,
\end{align}
where the nonrotating component $\psi_{g}(\mathbf r)$ and the rotating vortex components $\psi_{v\pm}({\mathbf r})$ are given by
\begin{align}
\psi_g({\mathbf r})&= \exp\{-({1}/{2})[({r}/{L_{\perp}})^2 + ({z}/{L_{z}})^2 ] \}/{\pi^{3/4} L_{\perp} L_{z}^{1/2}}\nonumber \\
 \psi_{v\pm}({\mathbf r})&={(x\pm \ri y)^{|\ell|}}/{\sqrt{|\ell |!} \, L_{\perp}^{|\ell |}}  \psi_g({\mathbf r})\forget{= {r^{|\ell|} e^{\pm \ri \ell \phi}}/{\sqrt{|\ell |!} \, L_{\perp}^{|\ell |} } \psi_g({\mathbf r})}\,.
\end{align}
\end{subequations}
Here $L_{\perp}$ and $L_{z}$ are the size parameters of the condensate in the $x$-$y$ plane and the $z$ directions respectively and $\delta$ is the two-photon detuning as shown in Fig.~\ref{Fig:LevelScheme}.
Thus, the time dependence of the populations of different components $(|\alpha(t)|^2, |\beta_{\pm}(t)|^2)$ can be studied by projecting the rate equations~\eqref{Eq:psidot} on to the topological states~\eqref{Eq:topology}. So far the equations are very general and no restriction exists on the OAM quantum number $\ell$. Hereafter, for convenience,  we resort to a particular value 
of $\ell=2$. However, one may note that the general idea would remain valid for any given $\ell$. Thus, taking the projections onto the specified rotating or non-rotating states we arrive at
\begin{align}
\ri \dot{\alpha}(t) &=  3 \kappa |\alpha(t)|^2 \alpha(t) + \omega_{\perp} \(a_{+}^* \beta_{+}(t) + a_{-}^* \beta_{-}(t)\)
\nonumber \\
\ri \dot{\beta}_{\pm}(t) &= (\delta  + 2 \omega_{\perp}+ 
\frac{\kappa}{2} \sum_{i=\pm}|\beta_{i}(t)|^2) \beta_{\pm}(t)
+ \omega_{\perp}\, a_{\pm}\, \alpha(t)
 \end{align}
 Here, the interparticle interaction strength appeares through the parameter $\kappa=\pi \hbar a_{\rm sc} N / [m (2 \pi)^{3/2} L_{\perp}^2 L_{z}]$.
This set of equations can be solved numerically, using the experimental parameters for a $^{87}$Rb BEC~\cite{Jin:1996} (i.e., $\omega_{\perp}=132 \text{ Hz}$, $a_{\rm sc} = 5 \text{ nm}$, $L_{\perp}=2.35\,\, \mu\text{m}$, $L_z=1.4\,\, \mu{\rm m}$), so that $\kappa= 422\,\, {\rm Hz}$. The results of our numerical studies are summarized in Figures~\ref{Fig:TF} and \ref{Fig:Pop}. We define a transfer function $f(t)= |\alpha(t)|^2-|\beta_{+}(t)|^2-|\beta_{-}(t)|^2$, which signifies the amount of population transferred from state $\ket{0}$ to states $\ket{\pm}$. Initially $f(t=0)=1$ as all the population  resides in the non-rotating ground state. If complete transfer is achieved to an appropriate superposition of the $\ket{\pm}$ states then $f(t)\rightarrow -1$. This transfer function is plotted for various values of the two-photon detuning $\delta$ in Fig.~\ref{Fig:TF}. The result to be noted is that only a continuous time variation of the detuning leads to complete population transfer at the steady state. This can be understood as the interaction terms in the dynamical equations effectively lead to a time-dependent stark shift of the different states, and only a time varying detuning maintains effective two-photon resonance to give complete population transfer. It can be shown that the final state obtained at steady state is $a_{+}^* \ket{+} + a_{-}^* \ket{-}$ where we started with the optical vortex state $a_{+}\ket{\ell} + a_{-} \ket{-\ell}$, i.e., the phase relation between the BEC vortex components is closely related  to the optical vortex components. In figure~\ref{Fig:Pop} we show generation of various superposition of the vortex states $\ket{\pm}$ for the time varying detuning.

These vortex states could be detected, for example, by following the proposal of Bolda and Walls~\cite{Bolda:1998a}, which involves observing a distinctive interference pattern of the vortex state with a non-rotating BEC cloud. However, such a measurement can not distinguish the sense of rotation the BEC cloud. We propose scattering of light incident on the cloud perpendicular to the rotation axis to observe a Doppler shift of the scattered light. Such a  measurement scheme would detect the sense of rotation of the vortex state and would collapse the state to either of the two counter-rotating components. We conjecture that such a coherent superposition of two counter-rotating vortex states would be ideally suited for gyropscopy and other forms of inertial sensing.  An entangled state of these vortices would also be interesting objects to study. Further discussions of these and related issues will be presented elsewhere.

\forget{\it Summary---}
To conclude, we have devised a scheme to generate superposition of two counter-rotating optical vortices. We have designed a Raman type scheme to transfer the optical vortex superposition onto a superposition of vortices in BEC. These macroscopic superpositions of two counter-rotating vortices are very interesting objects. We suggest a detection scheme based on rotational Doppler shift of a light field to sense the sense of rotation. 

\forget{\it Acknowledgments ---}
Part of this work, done by KTK, was carried out 
at the Jet Propulsion Laboratory under 
a contract with the National Aeronautics and Space Administration (NASA). 
KTK acknowledges support from the National Research Council and
NASA, Codes Y and S. KTK also wishes to thank Prof. K.-P. Marzlin for useful discussions. KTK and JPD wish to thank Prof. M. S. Zubairy for useful discussions. JPD acknowledges support from the Horace C. Hearne Jr. Foundation, the Advanced Project Research Agency, the National Security Agency, and the National Reconnaissance Office.


\begin{thebibliography}{0}
\expandafter\ifx\csname natexlab\endcsname\relax\def\natexlab#1{#1}\fi
\expandafter\ifx\csname bibnamefont\endcsname\relax
  \def\bibnamefont#1{#1}\fi
\expandafter\ifx\csname bibfnamefont\endcsname\relax
  \def\bibfnamefont#1{#1}\fi
\expandafter\ifx\csname citenamefont\endcsname\relax
  \def\citenamefont#1{#1}\fi
\expandafter\ifx\csname url\endcsname\relax
  \def\url#1{\texttt{#1}}\fi
\expandafter\ifx\csname urlprefix\endcsname\relax\def\urlprefix{URL }\fi
\providecommand{\bibinfo}[2]{#2}
\providecommand{\eprint}[2][]{\url{#2}}

\end{thebibliography}


\begin{thebibliography}{99}
\bibitem{Nielsen:2000}
M.~A. Nielsen and I.~L. Chuang, {\em Quantum Computation and Quantum
  Information} (Cambridge University Press, Cambridge, 2000).

\bibitem{Anderson:1995}
M.~H. Anderson {\it et~al.}, Science {\bf 269},  198  (1995).

\bibitem{Bradley:1995}
C.~C. Bradley, C.~A. Sackett, J.~J. Tollett, and R.~G. Hulet, Phys. Rev. Lett.
  {\bf 75},  1687  (1995).

\bibitem{Mewes:1996}
M.-O. Mewes {\it et~al.}, Phys. Rev. Lett. {\bf 77},  416  (1996).

\forget{\bibitem{Balykin:2000}
V.~I. Balykin, V.~G. Minogin, and V.~S. Letokhov, Rep. Prog. Phys. {\bf 63},
  1429  (2000).}

\bibitem{Abo-Shaeer:2001}
J.~R. Abo-Shaeer, C. Raman, J.~M. Vogels, and W. Ketterle, Science {\bf 292},
  476  (2001).

\bibitem{Heckenberg:1992}
N.~R. Heckenberg, R. McDuff, C.~P.~Smith, and A.~G.~White, Opt. Lett. {\bf
  17},  221  (1992).

\bibitem{Agarwal:1997OAM}
G.~S. Agarwal, R.~R. Puri, and R.~P. Singh, Phys. Rev. A {\bf 56},  4207
  (1997).

\bibitem{Arlt:1998}
J. Arlt, K. Dholakia, L. Allen, and M.~J. Padgett, J. Mod. Opt. {\bf 45},  1231
   (1998).

\bibitem{Sueda:2004}
K. Sueda, G. Miyaji, N. Miyanaga, and N. Nakatsuka, Opt. Exp. {\bf 12},  3538
  (2004).

\bibitem{Akamatsu:2003}
D. Akamatsu and M. Kozuma, Phys. Rev. A {\bf 67},  023803  (2003).

\bibitem{Molina-Terriza:2002}
G. Molina-Terriza, J.~P. Torres, and L. Torner, Phys. Rev. Lett. {\bf 88},
  013601  (2002).

\bibitem{Leach:2002}
J. Leach {\it et~al.}, Phys. Rev. Lett. {\bf 88},  257901  (2002).

\bibitem{Bigelow:2004}
M.~S. Bigelow, P. Zerom, and R.~W. Boyd, Phys. Rev. Lett. {\bf 92},  083902
  (2004).

\bibitem{Grier:2003}
D.~G. Grier, Nature (London) {\bf 424},  810  (2003).

\bibitem{Spedalieri:2004QKD}
F.~M. Spedalieri, quant-ph/0409057  (2004).

\bibitem{Mair:2001}
A. Mair, A. Vaziri, G. Weihs, and A. Zeilinger, Nature (London) {\bf 412},  313
   (2001).

\bibitem{Marzlin:1997}
K.-P. Marzlin, W. Zhang, and E.~M. Wright, Phys. Rev. Lett. {\bf 79},  4728
  (1997).

\bibitem{Nandi:2004}
G. Nandi, R. Walser, and W.~P. Schleich, Phys. Rev. A {\bf 69},  063606
  (2004).

\bibitem{Vaziri:2002}
A. Vaziri, G. Weihs, and A. Zeilinger, J. Opt. B {\bf 4},  S47  (2002).

\bibitem{Allen:1992}
R.~J. C.~S. L.~Allen, M. W.~Beijersbergen and J.~P. Woerdman, Phys. Rev. A {\bf
  45},  8185  (1992).

\bibitem{Padgett:1999}
M.~J. Padgett and J.~P. Lesso, J. of Mod. Opt. {\bf 46},  175  (1999).

\bibitem{Zeilinger:1981}
A. Zeilinger, Am. J. Phys. {\bf 49},  882  (1981).

\bibitem{Niu:2004}
Y. Niu, S. Gong, R. Li, and S. Jin, Phys. Rev. A {\bf 70},  023805  (2004).

\bibitem{Bergmann:1998}
K. Bergmann, H. Theuer, and B.~W. Shore, Rev. Mod. Phys. {\bf 70},  1003
  (1998).

\bibitem{Edwards:1996}
M. Edwards {\it et~al.}, Phys. Rev. A {\bf 53},  1950(R)  (1996).

\bibitem{Jin:1996}
D.~S. Jin {\it et~al.}, Phys. Rev. Lett. {\bf 77}, 420 (1996).

\bibitem{Bolda:1998a}
E.~L. Bolda and D.~F. Walls, Phys. Rev. Lett. {\bf 81},  5477  (1998).

\end{thebibliography}

\end{document}